\documentstyle[aps,prd,epsfig]{revtex}
\newcommand{\bga}{\mbox{\boldmath$\gamma$}}
\newcommand{\bSi}{\mbox{\boldmath$\Sigma$}}

\begin{document}
\draft

\title{Quasiquarks in two stream system}

\author{Stanis\l aw Mr\' owczy\' nski\footnote{Electronic address:
mrow@fuw.edu.pl}}

\address{So\l tan Institute for Nuclear Studies \\
ul. Ho\.za 69, PL - 00-681 Warsaw, Poland \\
and Institute of Physics, \'Swi\c etokrzyska Academy \\
ul. Konopnickiej 15, PL - 25-406 Kielce, Poland}

\date{25-th April 2002}

\maketitle

\begin{abstract}

We study the collective quark excitations in an extremely anisotropic 
system of two interpenetrating streams of the quark-gluon plasma. In 
contrast to the gluon modes, all quark ones appear to be stable in such 
a system. Even more, the quark modes in the two-stream system are 
very similar to those in the isotropic plasma.

\end{abstract}

\vspace{0.5cm}
%\pacs{PACS: 12.38.Mh, 11.10.Wx}


\vspace{0.5cm}

The quark-gluon plasma exhibits a rich spectrum of collective 
excitations. During the last two decades a lot of efforts has been 
paid to study the equilibrium plasma and the excitations are rather 
well understood in this case, see \cite{Bla01} for an extensive review. 
Much less is known about the collective modes in the nonequilibrium 
quark-gluon plasma such as the parton system generated at the early 
stage of  ultrarelativistic heavy ion collisions at RHIC or LHC. 
The parton momentum distribution is not istotropic but strongly elongated 
along the beam \cite{Gei95,Wan97}. Therefore, specific color fluctuations, 
instead of being damped, can exponentially grow and noticeably influence 
the temporal evolution of the whole system. In a series our papers 
\cite{Mro88,Mro93,Mro94} it has been argued that there are indeed very fast 
unstable transverse gluon modes in such a parton system. The analysis 
\cite{Mro93,Mro94} has been performed within the semiclassical transport 
theory \cite{Elz89,Mro89}. However, it has been later shown that one 
gets the equivalent results within the hard loop approach \cite{Bra90a} 
extended to the anisotropic systems \cite{Mro00}. In the same paper 
\cite{Mro00}, the quark self energy for an arbitrary momentum distribution 
has been computed at one loop level and the dispersion relation of quarks 
in the anisotropic plasma has been also briefly discussed. 

The aim of this note is to consider for the first time the quark modes in 
the anisotropic plasma. We start with an extremely anisotropic system of 
two interpenetrating parton streams. The gluon modes have been discussed 
in such a system in the several papers 
\cite{Mro88,Pok88,Pok90a,Pok90b,Pav92} 
and the instabilities have been found. We show here that in contrast to 
the gluon excitations all quasiquark modes in the two-stream system are 
stable and rather similar to those in the isotropic plasma. At the end 
we argue that our results obtained for two streams approximately hold  
for any system of strongly elongated momentum distribution.

The quark dispersion relations are determined by the poles of quark 
propagator or equivalently are found as solutions of the equation
\begin{equation}\label{dispersion-eq1}
{\rm det}\Big[ k\!\!\!/\,  - \Sigma (k^{\mu})  \Big]  = 0 \;,
\end{equation}
where $k^{\mu} = (\omega, {\bf k})$ is the mode's four momentum; 
the mass of the bare quark is assumed to vanish. The quark self 
energy $\Sigma (k^{\mu})$ has been found \cite{Mro00} within the 
Hard Loop approximation at one loop level in the following form:
\begin{equation} \label{self}
\Sigma (k^{\mu}) = \frac{g^2}{16}C_F \int \frac{d^3p}{(2\pi )^3}\:
\frac{f({\bf p})}{|{\bf p}| }\: 
\frac{\gamma _0 - {\bf v}\cdot \bga}
{\omega  - {\bf v}\cdot {\bf k}+i0^+}\; ,
\end{equation}
where $g$ is the coupling constant of QCD with $N_c$ colors and
$C_F \equiv (N_c^2 - 1)/N_c$; ${\bf v} \equiv {\bf p}/ |{\bf p}|$
is the parton velocity and $\gamma^{\mu}=(\gamma^0, \bga)$ are
Dirac matrices; 
$f({\bf p}) \equiv n({\bf p})  + \bar n({\bf p}) + 2 n_g({\bf p})$ 
and $n({\bf p})$, $\bar n({\bf p})$, and $n_g({\bf p})$ denote
distribution functions of, respectively, quarks, antiquarks and
gluons. The effective distribution function $f$ is arbitrary though
symmetric i.e. $f(-{\bf p}) = f({\bf p})$. 

Since the spinor structure of $\Sigma$ given by Eq.~(\ref{self})
is very simple i.e. 
$\Sigma (k^{\mu}) = \gamma^0 \Sigma^0(\omega,{\bf k}) 
- \bga \bSi (\omega,{\bf k})$, the dispersion equation 
(\ref{dispersion-eq1}) simplifies to 
\begin{equation}\label{dispersion-eq2}
\big( \omega - \Sigma^0(\omega, {\bf k}) \big)^2
- \big({\bf k} - \bSi (\omega, {\bf k}) \big)^2  =0 \;.
\end{equation}

Let us now consider an extremely anisotropic system of two interpenetrating 
streams with the distribution function of the form
\begin{equation}\label{distribution-f}
f({\bf p}) = (2\pi)^3 \rho \: [ \delta^{(3)}({\bf p}-{\bf q}) 
+ \delta^{(3)}({\bf p}+{\bf q}) ] \;,
\end{equation}
where the parameter $\rho$ is related to the parton density in the
stream. The vector ${\bf q}$ is assumed to be parallel to $z-$axis. 
Substituting (\ref{distribution-f}) into Eq.~(\ref{self}) one 
immediately finds the self energies
\begin{eqnarray}\label{self2}
\Sigma^0(\omega, {\bf k}) = m^2 {\omega \over \omega^2 - k_z^2}\;, 
\\ [2mm] \nonumber
\Sigma^x(\omega, {\bf k})  = \Sigma^y(\omega, {\bf k}) = 0 \;,
\;\;\;\;\;\;\;\;\;
\Sigma^z(\omega, {\bf k})  = m^2 {k_z \over \omega^2 - k_z^2} \;,
\nonumber
\end{eqnarray}
where $m^2 \equiv g^2 C_F \rho /(8|{\bf q}|)$, and gets the following 
dispersion equation:
\begin{equation}\label{dispersion-eq3}
{\omega^2 (\omega^2 - k_z^2 - m^2)^2 \over (\omega^2 - k_z^2)^2}
- k_T^2
-{k_z^2 (\omega^2 - k_z^2 - m^2)^2 \over (\omega^2 - k_z^2)^2} = 0 \;,
\end{equation}
with $k_T \equiv \sqrt{k_x^2 + k_y^2}$. Eq.~(\ref{dispersion-eq3}) 
is easily solved providing two quark ($\omega_{\pm} > 0$) and 
two antiquark ($\omega_{\pm} < 0$) modes:
\begin{equation}\label{modes-2s}
\omega_{\pm}^2(k_T,k_z) = m^2 + k_z^2 + {1 \over 2} k_T^2
\pm \sqrt{m^2 k_T^2 + {1\over 4}k_T^4 } 
\cong m^2 + k_z^2 + {1 \over 2} k_T^2 \pm m \, k_T \;,
\end{equation}
where the second approximate equality holds for $m \gg k_T$.
Since the modes are pure real, there is no instability.
The dispersion relations (\ref{modes-2s}) are illustrated in
Figs. 1 and 2.

\vspace{-0.5cm}
\begin{figure}
\centerline{\epsfig{file=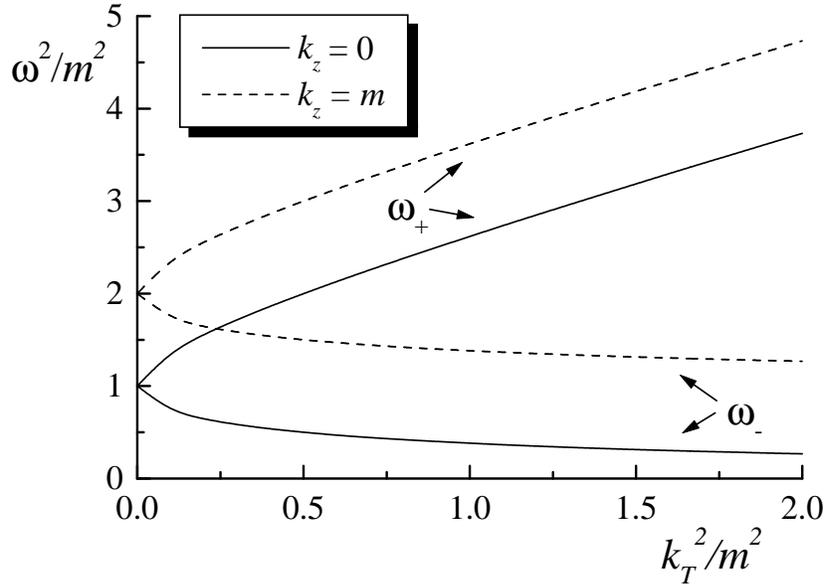,bbllx=14pt,bblly=440pt,bburx=540pt,bbury=850pt,width=0.6\linewidth}}
\vspace{0.9cm}
\caption{$\omega^2_{\pm}$ as a function of $k_T$ for
$k_z = 0$ and $k_z = m$.}
\end{figure}

We are going to compare now the quark modes in the two-stream system 
(\ref{modes-2s}) to those in the isotropic plasma studied, in 
particular, in \cite{Wel82,Bla93}. In contrast to the derivations
in \cite{Wel82,Bla93}, we do not assume that the system is in the
thermodynamic equilibrium. We only demand the isotropy of the
parton momentum distribution i.e. the distribution function which
enters the self energy (\ref{self}) is of the form $f({\bf p})
= f(|{\bf p}|)$. Then, the angular and momentum integrals in 
Eq.~(\ref{self}) factorize and one finds
\begin{eqnarray}\label{self0-eq}
\Sigma^0(\omega, {\bf k}) &=& { M^2 \over 2k} \Big[ \,
{\rm ln}\Big| {\omega + k \over \omega -k} \Big| 
-i \pi \Theta (k - \omega) \Big] 
= {M^2 \over \omega} \big( 1 + {\cal O}(k^2/\omega^2) \big) \;, \\ [3mm]
\label{selfv-eq}
\bSi (\omega,{\bf k}) &=& { M^2 \over k^2} \; {\bf k}
- { M^2 \omega \over 2k^3 } \Big[ \,
{\rm ln}\Big| {\omega + k \over \omega -k} \Big| 
-i \pi \Theta (k - \omega) \Big] \; {\bf k} 
= {\cal O}(k^2/\omega^2) \; { M^2 \over k^2} \; {\bf k} \;,
\end{eqnarray}
where $k \equiv |{\bf k}|$ and
$$
M^2 \equiv \frac{g^2}{32 \pi^2}C_F \int_0^{\infty} dp p \, f(p) \;.
$$
In the case of massless baryonless plasma in thermal equilibrium $M$ is
proportional to the system's temperature.

\vspace{-0.5cm}
\begin{figure}
\centerline{\epsfig{file=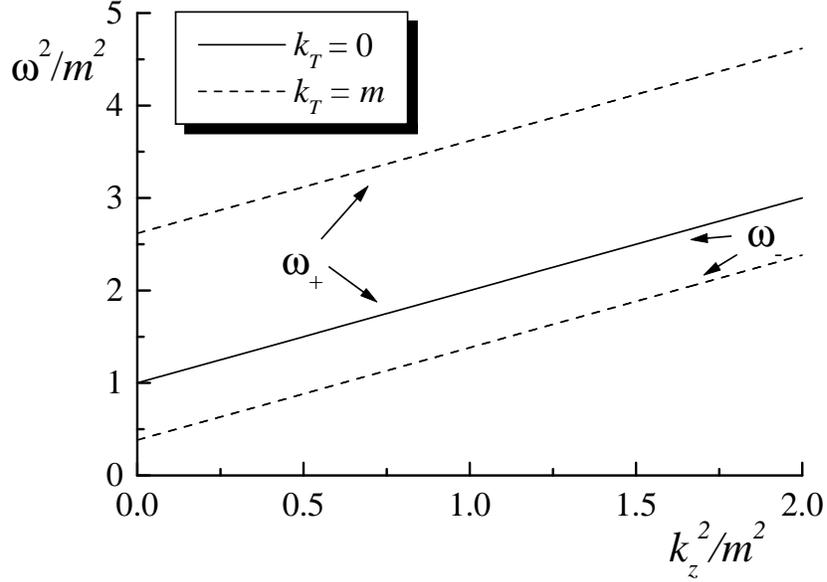,bbllx=14pt,bblly=440pt,bburx=540pt,bbury=85
0pt,width=0.6\linewidth}}
\vspace{0.9cm}
\caption{$\omega^2_{\pm}$ as a function of $k_z$ for $k_T = 0$
and $k_T = m$.}
\end{figure}

Substituting the self energies (\ref{self0-eq},\ref{selfv-eq})
into the dispersion equation (\ref{dispersion-eq2}), one finds
the quark modes which in the long wavelength limit ($ \omega \gg k$)
can be computed analytically and read
\begin{equation}\label{modes-eq}
\omega_{\pm}^2(k) \cong M^2 + {1 \over 2} k^2 \pm M \, k \;.
\end{equation}
As seen, the quark long wavelength modes in the two-stream system 
(\ref{modes-2s}) and in the isotropic plasma (\ref{modes-eq}) are very 
similar to each other. The $\omega_+$ mode is nearly 
the same in the two systems for any wavelength. However, there is 
a qualitative difference in the case of the $\omega_-$ mode. 
In equilibrium, $\omega_-$ initially decreases with $k$ according 
to Eq.~(\ref{modes-eq}), then there is a shallow minimum at 
$k \approx M$ and further $\omega_-$ monotonically grows. 
In the two-stream system, $\omega_-$ is monotonically decreasing
function of $k_T$. Since the phase velocity of the $\omega_-$ 
mode is greater than the velocity of light for sufficiently large 
$k_T$, the quasiquarks can emit gluons due to the Cherenkov mechanism.
The relevance of the phenomenon for the nonequilibrium plasma
from the early stage of heavy-ion collisions at RHIC and LHC
is under study.

One observes that the form of the quark self energy (\ref{self2}) 
and consequently the dispersion relations (\ref{modes-2s}) hold not
only for a highly simplified distribution function (\ref{distribution-f}) 
but for any strongly elongated momentum distribution.  Indeed,
if $\langle p_z^2 \rangle \gg \langle p_T^2 \rangle$ the velocities
under the integral (\ref{self}) can be approximated as $v_z \cong 1$
and $v_{x,y} \cong 0$. Then, one immediately gets the formulas
(\ref{self2})  with 
$$
m^2  = \frac{g^2}{16}C_F \int \frac{d^3p}{(2\pi )^3}\:
\frac{f({\bf p})}{|{\bf p}| } \;.
$$

We conclude our considerations as follows. In contrast to the 
gluon modes, the quark ones are all stable in the two-stream 
system. In the long wavelength limit, the quasi quark properties 
in the strongly anisotropic and isotropic systems are qualitatively 
the same. The $\omega_-$ mode can be responsible for the Cherenkov
radiation.

\end{document}